\documentclass[aps,a4paper,showpacs,hyphens]{revtex4}
\usepackage{epsfig}
\usepackage{graphicx}
\usepackage{amsmath,amssymb,color}
\usepackage[english]{babel}
\usepackage[colorlinks=true, allcolors=blue]{hyperref}
\usepackage{comment}

\parskip=\medskipamount

\newcommand{\eq}[1]{(\ref{#1})}
\newcommand{\fig}[1]{Fig. \ref{#1}}

\newcommand{\be}{\begin{equation}}
\newcommand{\ee}{\end{equation}}

\newcommand{\barr}{\begin{array}}
\newcommand{\earr}{\end{array}}

\newcommand{\beqn}{\begin{eqnarray}}
\newcommand{\eeqn}{\end{eqnarray}}

\newcommand{\bs}{\begin{subequations}}
\newcommand{\es}{\end{subequations}}

\newcommand{\bw}{\begin{widetext}}
\newcommand{\ew}{\end{widetext}}

\newcommand{\Nint}{N_\mathrm{int}} 

\definecolor{green1}{rgb}{0.1, 0.6, 0.1} 

\DeclareGraphicsExtensions{.png,.pdf}

\begin{document}

\title{Learning thresholds lead to stable language coexistence}

\author{M.V.~Tamm$^1$, E.~Heinsalu$^2$, S.~Scialla$^{2,3}$, M.~Patriarca$^2$}

\affiliation{$^1$ School of Digital Technologies, Tallinn University, Tallinn, Estonia; \\
$^2$  National Institute of Chemical Physics and Biophysics, Tallinn, Estonia; \\
$^3$ Università Campus Bio-Medico di Roma, Rome, Italy}

\date{\today}

\begin{abstract}
We introduce a language competition model that is based on the Abrams-Strogatz model and incorporates the effects of memory and learning in the language shift dynamics. 
On a coarse grained time scale, the effects of memory and learning can be expressed as thresholds on the speakers fractions of the competing languages.
In its simplest form, the resulting model is exactly solvable.
Besides the consensus on one of the two languages, 
the model describes additional equilibrium states that are not present in the Abrams-Strogatz model: a stable dynamical coexistence of the two languages and a frozen state coinciding with the initial state.
We show numerically that these results are preserved for threshold functions of a more general shape.  
The comparison of the model predictions with historical datasets demonstrates that while the Abrams-Strogatz model fails to describe some relevant language competition situations, the proposed model provides a good fitting. 
\end{abstract}

\maketitle

\textit{Introduction}. --
In multilingual societies languages are in constant competition with each other. 
This competition manifests itself in language shift, meaning that individuals switch the language they use, and can lead to a gradual change of the linguistic composition of the society and even to the eventual disappearance of one of the languages. 

Starting from the seminal papers of Baggs and Freedman \cite{Baggs-1990a} and Abrams and Strogatz \cite{Abrams-2003a}, language competition has been studied through models mathematically similar to the ones describing ecological competition or chemical reaction kinetics.
Various generalizations of these models have been proposed, in order to take into account additional aspects of the underlying language dynamics and social processes \cite{Patriarca-2012a,Patriarca-2020a}, e.g., the presence of bilinguals~\cite{Minett2008a,Baronchelli-2006a,Heinsalu-2014a},
the effects of political and geographic barriers~\cite{Patriarca-2004a,Patriarca-2009a}, 
population dynamics~\cite{Baggs-1993a,Pinasco-2006a,Kandler2008a}, 
the inhomogeneity of resources~\cite{Kandler2009a}, and the similarity between languages~\cite{Mira2005a}. 
These models have been used to study also other aspects of social dynamics, such as change in religious affiliation \cite{Abrams2011} and political polarization \cite{Lu2019}, thus making language competition an archetypal problem of social dynamics. 

The mean-field models, such as the Abrams-Strogatz (AS) model, usually predict that the outcome of the language competition is the extinction of one of the competing languages. 
Thus, they fail to explain the slowing down of language decay and the coexistence of competing languages that is often observed. 
Here we suggest a generalized AS model, which solves this deficiency and suggests a possible microscopic mechanism behind long-term language coexistence. 

We show that the language coexistence can be explained taking into account 
the language learning process, which is a prerequisite for language shift \cite{Scialla-2023a} and suggests the existence of thresholds in the language shift rates. 
Namely, 
It is known since Ebbinghaus' study of human memory that 
forgetting can be reduced and the amount of information retained in the long-term can be increased through sufficiently frequent repetitions \cite{Ebbinghaus-1885a,Murre-2015a}. 
Therefore in order to learn a language organically, a sufficiently high rate of encounters with speakers of this language is needed.
This in turn translates into a threshold on the corresponding fraction of speakers. 

Alternatively, learning a new language may not happen organically through encounters with speakers of the language, but it may be the result of a conscious decision of an individual to allocate time specifically for learning. 
For such a conscious learning to be feasible, an individual should perceive the time spent to learn the language as beneficial compared to other activities and thus it can be described as an outcome of social pressure. 
In various studies of social dynamics \cite{Granovetter-1978}, opinion dynamics \cite{Liggett-1994, Vieira-2018}, learning \cite{Gonzalez-2011}, and contagion \cite{Contreras-2024}, the social pressure has been described through threshold models. 
Therefore, it is expected that also the dynamics of language shift is induced by the fraction of speakers of a language exceeding a certain threshold.



\begin{figure}
    \centering
    \includegraphics[width=8.5cm]{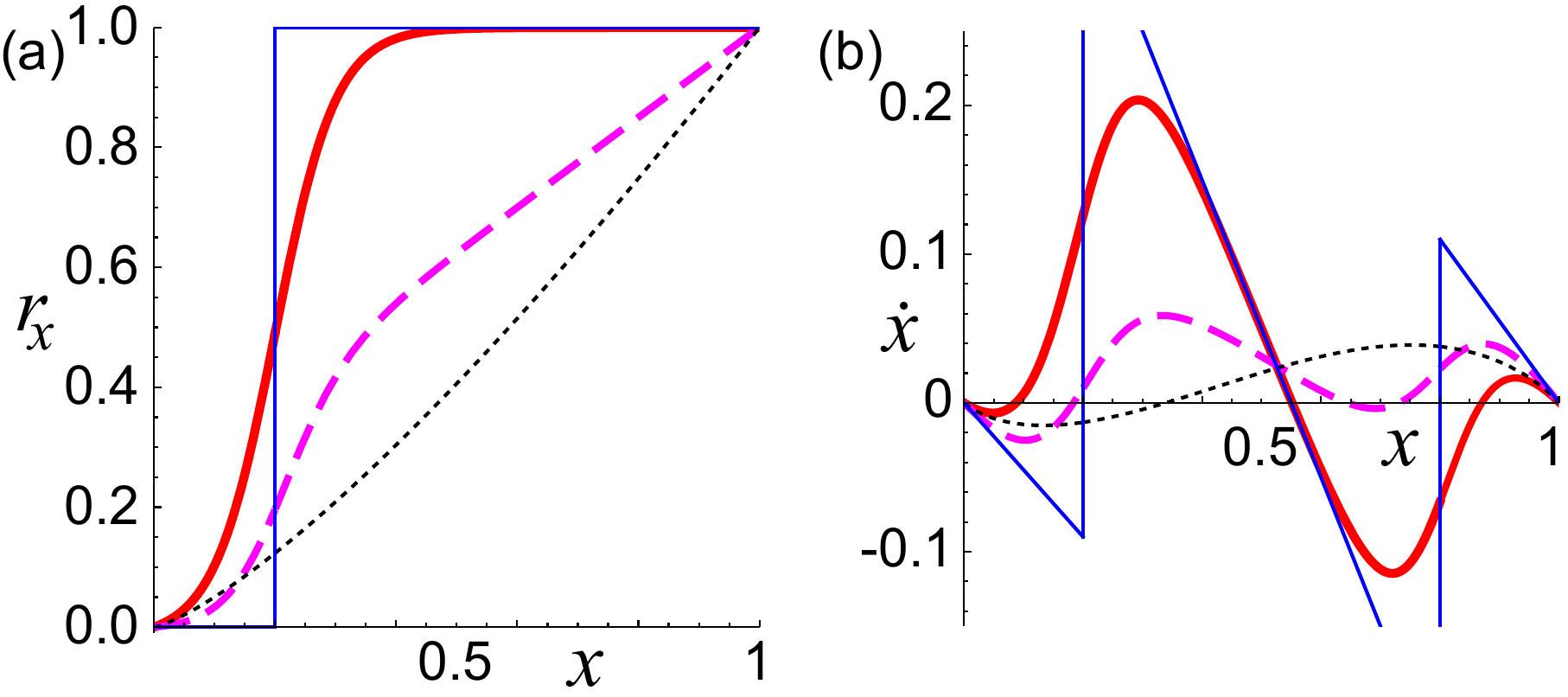}
    \caption{
    Panel~(a): Transition rates $r_x(x)$. 
    Black dotted line -- power law rate \eqref{rates} of the AS model with $a = 1.3$; 
    thin blue line -- step function rate defined by Eqs.~\eqref{step}-\eqref{const}; 
    thick red line -- generalized sigmoid rate \eq{SShape_replacement} with $w = 0.1$ and constant $r_{0x}$; 
    dashed magenta line -- generalized sigmoid rate with $w = 0.1$ and $r_{0x}(x) = 0.25 + 0.75x$.
    For all transition rates $j_x = 1, x^* = 0.2$. 
    Panel~(b): The $x$-velocity field defined by Eq.~\eqref{AS0}, computed for the same transition rates as in panel~(a). 
    We have assumed $y^* = x^*, \gamma = 0.55$. }
    \label{fig_rates}
\end{figure}

\textit{Abrams-Strogatz model.} -- 
The AS model \cite{Abrams-2003a} is a two-state language competition model, where individuals speak either language X or Y and can at any time undergo a language shift.  
The fractions of the X- and Y-speakers, $x(t)$ and $y(t)$, satisfy the condition $x(t) + y(t) = 1$ and the model can be described by the following mean-field dynamical equation: 
\begin{equation}
\label{AS0}
    \begin{array}{rll}
     \dot{x} = y r_{x}(x)  - x r_{y}(y)  = (1 - x) r_{x}(x)  - x r_{y}(1 - x) \, .
     \end{array}
\end{equation}
The transition rates $r_{x}$ for Y $\to$ X and $r_y$ for X $\to$ Y are assumed to be monotonously increasing functions of their arguments. 
The boundary conditions $r_{x}(x\!\to\!0) = 0$ and $r_y(y \to 0) = 0$ express the fact that language shift does not take place in the absence of speakers of the target language. 
In the AS model \cite{Abrams-2003a}
\begin{equation}
\begin{aligned}
\label{rates}
    r_{x}(x) = j_x x^a \, , \quad
    r_{y}(y) = j_y y^a \, , 
\end{aligned}
\end{equation}
where the exponent $a > 0$ is called volatility \cite{Castello2006a} and the parameters $j_{x}$, $j_{y}$ represent the maximal rates of language shift, reached when a speaker is immersed in a linguistic environment composed solely of speakers of the target language. 
Notice that in analogy with reaction kinetics models, e.g., the language shift $\mathrm{Y} \to \mathrm{X}$ would represent a chemical reaction $\mathrm{X} + \mathrm{Y} \to 2 \mathrm{X}$. 
This would suggest that $a = 1$ in the expression for $r_x$. 
However, analysis of real data shows that $a \in [1, 2]$ \cite{Abrams-2003a,Isern-2014a}.
The power law transition rate $r_x(x)$ of the AS model is illustrated in Fig.~\ref{fig_rates}(a) for $a = 1.3$ (black dotted line).

\begin{figure}
    \centering
    \includegraphics[width=8.5cm]{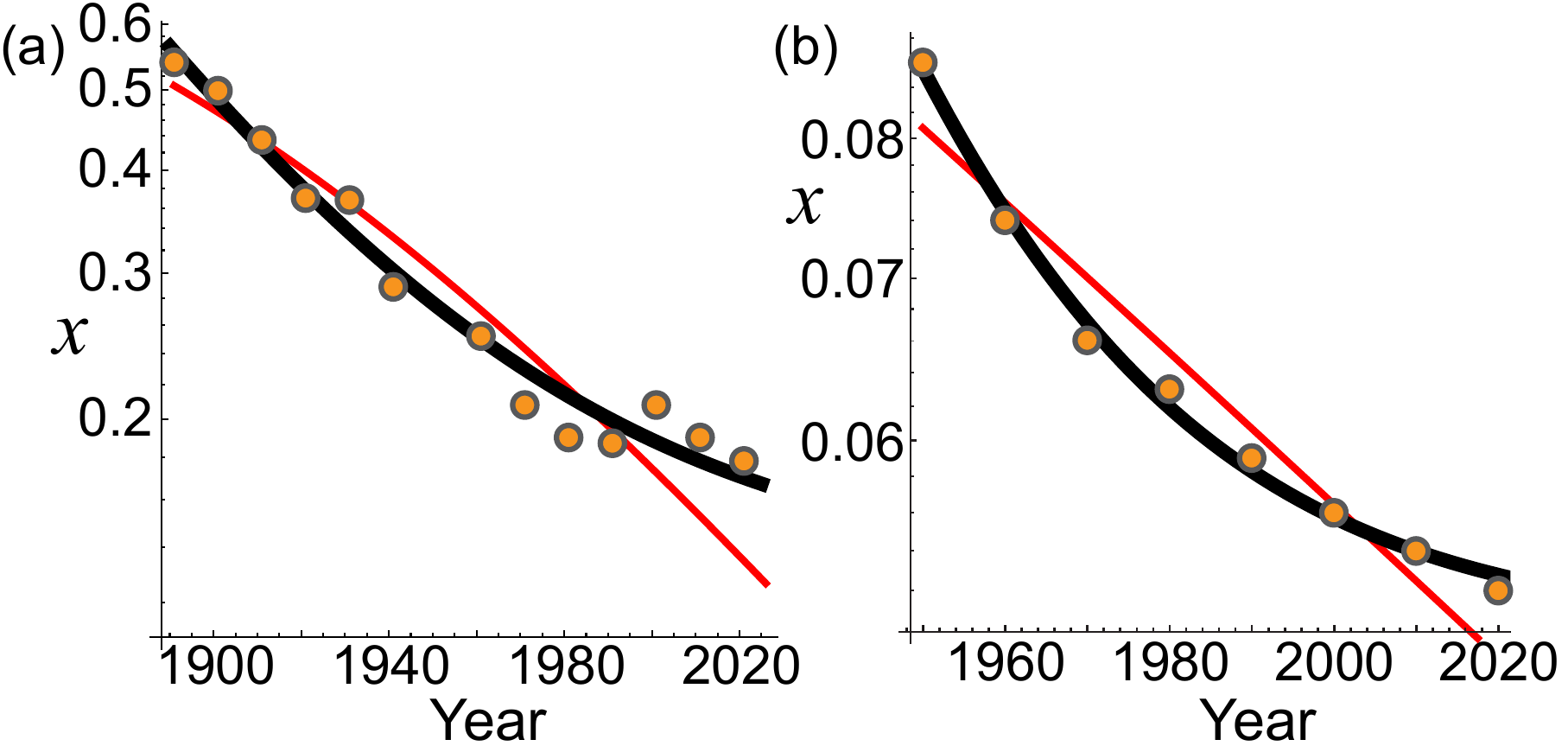}
    \caption{
    Semi-logarithmic plot of the fraction of speakers of Welsh in Wales [panel~(a)] and Swedish in Finland [panel~(b)]. 
    The points corresponding to census data \cite{wales,finland} are well fitted by the solution \eqref{coexponent} of the proposed model (thick black lines). 
    Thin red lines correspond to the best fits of the AS model with $a \geq 1$. 
    See Appendix C3 for the details of the fits .}
    \label{fig_data}
\end{figure}

Let us measure time in units of $(j_x + j_y)^{-1}$, which is observed to be around $50$ years, supported by the fact that migrants typically lose their original language in three generations. 
Then, introducing the dimensionless rate $\gamma = j_x/(j_x + j_y)$, that quantifies the prestige of language X and reflects the social effects and cultural values that influence the language choice, Eq.~(\ref{AS0}) becomes
\begin{equation}
\label{AS1}
    \dot{x} =  \gamma (1 - x) x^a  - (1 - \gamma) x (1 - x)^a  \, .
\end{equation}
The long-time behavior of the solution of this equation depends on the values of $a$ and $\gamma$.
For $a > 1$ there are two attractive equilibrium points, $x = 0$ and $x = 1$. 
For $a = 1$ there is a single attractive point: $x = 1$ for $\gamma>1/2$ and $x = 0$ for $\gamma < 1/2$.
Importantly, Eq.~\eq{AS1} implies that the rate of relative decline of a language always increases when the number of speakers decreases (see Appendix A1 for details), i.e., if  $dx/dt < 0 $ then 
\begin{equation} \label{concave}
    \frac{d^2}{dt^2}\left(\log x \right) 
=   \frac{d}{dt}\left(\frac{1}{x}\frac{dx}{dt}\right)
    < 0 \, .
\end{equation} 
%
However, as shown in Fig.~\ref{fig_data}, where two semi-logarithmic plots of the fraction of speakers are depicted, this is often not the case. 
In these examples, the decline of the minority language is clearly slowing down (the curve is convex), contrary to the prediction (\ref{concave}) of the AS model. 
Moreover, in other examples (see Appendix C2) the rate of language shift is extremely slow, of the order $10^{-3}$ per year, implying that the language composition is almost frozen.


%
%

\textit{Abrams-Strogatz model with step-function learning rates.} -- 
Learning a language is a prerequisite for language shift. 
For the reasons outlined in the introduction, we expect the rates $r_x$, $r_y$ to have a shape different from Eq.~(\ref{rates}), assumed in the AS model. 
Indeed, we expect that an individual is able to learn a language only if the interaction rate with the speakers of this language exceeds some critical value.
If all X-speakers are identical, one expects the existence of a critical concentration $y^*$, below which language Y cannot be learned, and an analogous critical concentration $x^*$ for Y-speakers, suggesting a rate function of the form [thin blue line in Fig.~\ref{fig_rates}(a)]
\begin{equation}
    r_x = r_{0x}(x) \Theta(x-x^*) \, , \quad r_y = r_{0y}(y) \Theta(y-y^*) \, .
    \label{step}
\end{equation}
%
The thresholds $x^*, y^*$ can in general be different from each other, reflecting the differences in language learning difficulty or prestige.
The functions $r_{0x}(x)$ and $r_{0y}(y)$ are positive and non-decreasing, 
satisfying the conditions $r_{0x}(x\!=\!1) = j_{x}$ and $r_{0y}(y\!=\!1) = j_{y}$. 

In the simple case of constant rates,
\begin{equation}
    r_{0x}\equiv j_x \, , \quad r_{0y}\equiv j_y \, ,
    \label{const}
\end{equation} 
Eq.~\eqref{AS0} becomes
\begin{equation}
\label{gAS1}
    \dot{x}  = \gamma \Theta(x - x^*) \, (1-x) - (1-\gamma) \Theta(1-x - y^*)  \, x \, .
\end{equation}
The dynamical evolution described by Eq.~\eqref{gAS1} is controlled by the step functions $\Theta(x-x^*)$ and $\Theta(1-x-y^*)$. 
For each pair of parameters $x^*, y^*$, the conditions $x=x^*$ and $1-x =y^*$ 
generate four possible options, corresponding to the two $\Theta$-functions assuming a value equal to zero or one.
For each of these options, Eq.~\eqref{gAS1} becomes linear and can be solved exactly, unlike the AS model.

If $x \in (1 - y^*, x^*)$, i.e., the arguments of both step functions are negative, Eq.~\eqref{gAS1} reduces to  $\dot{x} = 0$, and $x(t) \equiv x_0 ~ \forall t$. 
This frozen regime describes a population where interaction rates are so low that nobody will learn the other language and undergo language shift. 

If $x > \max (x^*, 1 - y^*)$, then $\Theta(x - x^*) = 1$ and $ \Theta(1 - x - y^*) = 0$ and Eq.~\eqref{gAS1} reduces to $\dot{x} = \gamma (1 - x)$. 
The solution of this equation increases with time, $x = 1 - (1 - x_0)\exp (-\gamma t)$, and therefore, if the condition $x > \max (x^*, 1-y^*)$ is satisfied initially, it remains satisfied at all subsequent moments. 
This regime corresponds to a system containing enough X-speakers, so that Y-speakers will learn language X, but not vice versa, and the dynamics eventually converges to consensus in language X, i.e., to solution $x_{\text{I}} = 1$.

The case $x < \min (x^*, 1-y^*)$ is analogous up to replacement $x \leftrightarrow 1 - x, \gamma \leftrightarrow 1 - \gamma$ and leads to the convergence to consensus in language Y, corresponding to the attractive stable point $x_{\text{II}} = 0$.

\begin{figure}
\centering
\includegraphics[width=9cm]{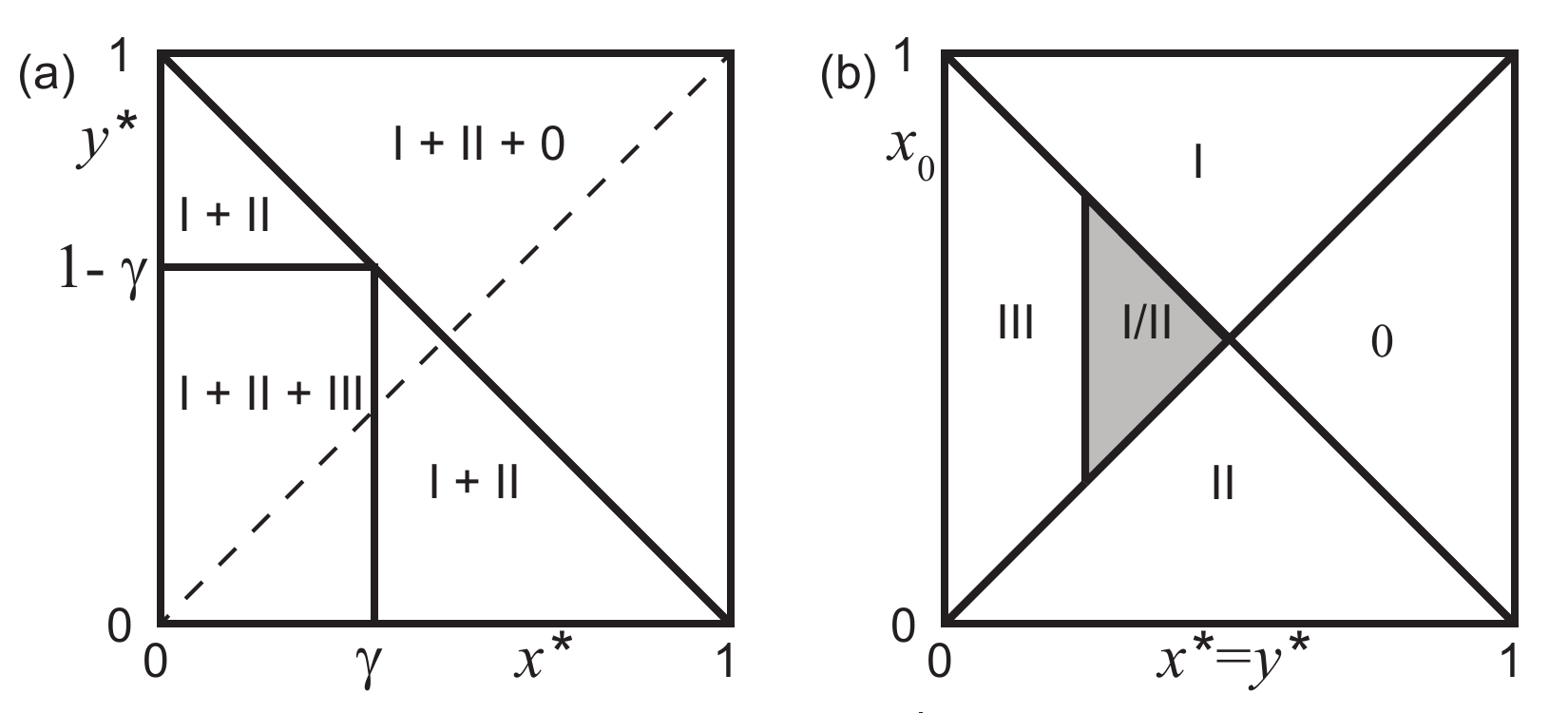}    
\caption{
    \label{Diagram}
    Panel (a):  
    Plane $x^*$-$y^*$ split into different regions corresponding to possible outcomes of the dynamics defined by Eq.~\eqref{gAS1}: 
    state 0 represents a dynamics frozen at the initial condition $x_0 = x(0)$; 
    state I corresponds to the X-consensus attractive fixed point $x_\text{I} = 1$; 
    state II -- to Y-consensus point $x_\text{II} = 0$;
    state III -- to the coexistence point $x_\text{III} = \gamma$. 
    Panel (b): $x_0$ vs $x^*(y^*)$ cross-section of the $x_0$-$x^*$-$y^*$ space along the dashed line $x^* = y^*$ in panel (a), showing the dependence on the initial conditions $x = x(0)$: each region leads to the unique asymptotic solution indicated, apart from the shaded region, defined by $x^* = y^* = \min(\gamma, 1 - \gamma)$, $x^* < x_0 < 1 - x^*$, leading to $x_\text{II} = 0$ if $\gamma < 1/2$ or $x_\text{I} = 1$ if $\gamma > 1/2$. 
}
\end{figure}

Finally, if $x \in [x^*, 1 - y^*]$, both step functions are equal to one, i.e., there are enough contacts in the system, so that both transitions from X to Y and from Y to X are possible.
In this case, Eq.~\eqref{gAS1} becomes  $\dot{x} = \gamma - x$, with the solution 
\begin{equation}
    x(t) = \gamma - (\gamma - x_0)\exp(-t)
    \label{coexponent}
\end{equation} 
monotonously approaching the limiting value $x_{\text{III}} = \gamma$. 
Importantly, contrary to the previous cases, the constraint $x(t) \in (x^*, 1 - y^*)$ defining this dynamical regime may not be fulfilled for all $t > 0$. 
If $\gamma \in [x^*, 1 - y^*]$, the dynamics remains in the same region and the solution converges to the attractive stable point $x = x_{\text{III}}$.
If $\gamma$ is above the upper limit of the interval, $\gamma > 1-y^*$, the solution $x(t)$ will increase until it reaches the region $x > \max (x^*, 1 - y^*)$ at some time and at long times gets attracted towards the stable point $x = x_{\text{I}}$. 
If $\gamma$ is below the lower limit of the interval, $\gamma < x^*$, then at some time the solution will reach the region $x < \min (x^*, 1 - y^*)$ and will eventually converge towards the stable point $x = x_{\text{II}}$.

\begin{figure*}
\centering
\includegraphics[width=17cm]{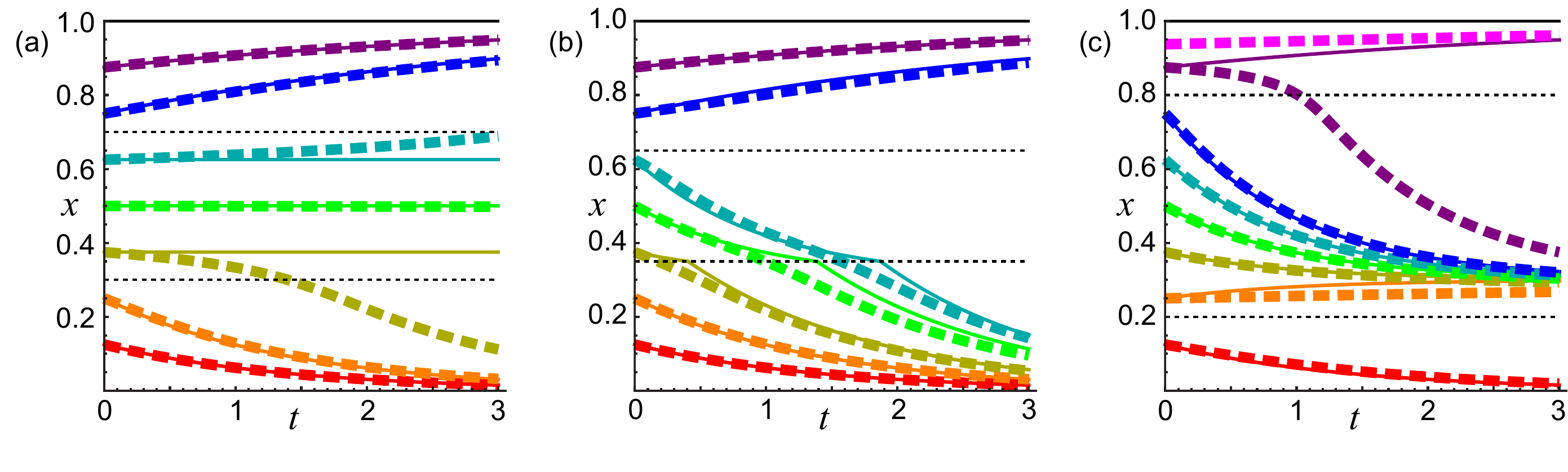}    
\caption{
    \label{dynamics}
    Time-evolution of the fraction of X-speakers for different initial conditions $x(0) = 1/8, 2/8, \dots, 7/8$ (from red to purple) and thresholds $x^* = y^*$; $\gamma = 0.3$.
    Solid lines correspond to the step-function rates, dashed lines to the sigmoid rates \eqref{SShape_replacement} with $w=0.07$. 
     Panel (a): $x^* = 0.7$; possible outcomes are I, II and 0 [see Fig.~\ref{Diagram}(b)]; for sigmoid rates the 0-regime becomes unstable and $x$ converges to either $x_{\text{I}}$ or $x_{\text{II}}$. 
    Panel (b): $x^* = 0.35$, possible outcomes are I, II; for step-function rates, there is a discontinuity in $\dot{x}$ when crossing the value $x = x^*$.
    Panel (c): $x^* = 0.2$, possible outcomes are I, II and III. 
    Using sigmoid rates, the boundaries of the basins of attraction move: e.g., in panel (c) the initial condition $x_0 = 7/8$ is attracted to the intermediate stable point  $x_{\text{III}}$; 
    however, outcome I is still possible, as illustrated by trajectory with $x_0 = 15/16$ (magenta dashed line).
}
\end{figure*}

The possible outcomes of the dynamics are summarized by Fig.~\ref{Diagram}.
Figure~\ref{Diagram}(a) shows the $x^*$-$y^*$ state diagram, split into different regions depending on the possible solutions,  indicated in the same area; e.g., in the region labeled ``0 + I + II'', the system can remain frozen in the initial state (0) or converge to the consensus in X (I) or in Y (II).
Figure~\ref{Diagram}(b) illustrates the dependence of the outcome on the initial condition $x_0 = x(0)$ in the case $x^* = y^*$. 
The plane is divided into disjoint regions corresponding to different outcomes.
If the initial conditions of the system are within the gray region, the system will converge either to state I of X-consensus, if $\gamma > 1/2$, or to state II of Y-consensus, if $\gamma < 1/2$.

In Fig.~\ref{dynamics} we plot with solid lines $x(t)$ for $\gamma = 0.3$ and three threshold values of $x^* = y^*$.
For $x^* = 0.7$ [panel (a)], initial conditions $x_0 > 0.7$ and $x_0 < 0.3$ lead to the attracting fixed points
$x_{\text{I}} = 1$ and $x_{\text{II}} = 0$, respectively, while for intermediate initial conditions the dynamics remains frozen. 
If $x^* = 0.35$ [panel (b)], the basins of attraction of $x_{\text{I}}$ and $x_{\text{II}}$ are separated by $x_0 = 1 - y^* = 0.65$. 
Note that trajectories starting	between $x^*$ and $y^*$ consist of two different exponential parts; the kink corresponds to $x = x^*$, at which the first step function in Eq.~\eqref{gAS1} changes its value. 
Finally, if $x^* = 0.2$ [panel (c)], initial conditions  $x_0 > 0.8$ and $x_0 < 0.2$ are still attracted to  $x_{\text{I}}$ and $x_{\text{II}}$, respectively, but the intermediate initial conditions converge to $x_{\text{III}} = \gamma = 0.3$, corresponding to a stable dynamic equilibrium between languages.

\textit{Abrams-Strogatz model with sigmoid learning rates.} -- 
Because people are not identical in their linguistic traits,  they should be characterized by different threshold values $x^*$, $y^*$.
If the distribution of these traits is relatively narrow, this changes the step-function $\Theta$ in Eq.~\eqref{step} into a sigmoid function $H$, here chosen as 
\begin{equation}
\begin{array}{rll}
    H(x,x^*,w) &=& \displaystyle \frac{S(x,x^*,w)-S(0,x^*,w)}{S(1,x^*,w)-S(0,x^*,w)} \, , \medskip \\
    S(x,x^*,w) &= & \{1+ \tanh[(x-x^*)/w]\}/2 \, ,
\end{array}
\label{SShape_replacement}
\end{equation}
which satisfies the boundary conditions $H(0, x^*, w) = 0$, $H(1, x^*, w) = 1$, and converges to $\Theta(x - x^*)$ for $w\to 0$ (see Appendix B3 for the details)

From Fig.~\ref{dynamics}, it can be observed that when the step-function is replaced by the sigmoid function, the neutral equilibrium due to frozen dynamics becomes unstable and the trajectory slowly converges either to $x_\text{I} = 1$ or $x_\text{II} = 0$.

\textit{Mechanism behind the language coexistence.} -- 
The model \eqref{gAS1} exhibits two additional regimes not present in the AS model: 
a frozen state for $x^* + y^* > 1$ 
and an attractive stable point corresponding to the coexistence of languages X and Y for $x^* + y^* < 1$. 

The existence of a frozen state follows from the assumption that language shift does not occur if the interaction rate among individuals is below some universal threshold.
The replacement of the step function with the sigmoid shape describes a more realistic scenario, where language shift rates are not identically zero but strongly suppressed if the interaction rates are below the threshold   (corresponding to fractions of speakers $x < x^*$ or $y < y^*$), e.g., the well-known lack of interactions between the Dutch- and French-speaking communities in Belgium~\cite{Blondel2008}. 

In turn, language coexistence for $x^* + y^* < 1$ is not due to a freezing of the dynamics but to a dynamical equilibrium, in which the same number of speakers switch language with the same rate in the two opposite directions, X $\to$ Y and Y $\to$ X. 
Figure~\ref{fig_rates}(a) compares the shapes of various Y $\to$ X shift rates $r_x(x)$.
%
%
The corresponding velocity field $\dot{x}$ as a function of $x$ associated to the different rates, see Fig.~\ref{fig_rates}(b), demonstrate some fundamental differences not revealed by the rates in panel~(a).
A point $x \in (0, 1)$ is an attractive stable point only if the velocity field is such that
$\dot{x} = 0$ and  $\partial \dot{x}/\partial x < 0$ \cite{Strogatz-2018a}.
Figure~\ref{fig_rates}(b) shows that there is no such point for the AS model for $a \geq 1$, whereas it exists for all the other models considered. 
Thus, the absence of language coexistence in the AS model is an artifact of the chosen shape of rate functions and even a relatively small change of the power-law rate form -- reflecting an underlying learning process -- can lead to stabilization of a state where two competing languages coexist. 

In Fig.~\ref{fig_data} we show the historical census data for the fraction of speakers of Welsh in Wales~\cite{wales} and Swedish in Finland~\cite{finland}, providing evidence of convergence to stable language coexistence. 
The convergence to equilibrium is exponential, as predicted by Eq.~\eqref{coexponent}.

\textit{Conclusion and outlook.} --
In this letter we present a language competition model that takes into account the dynamics of the language learning process and allows an analytical investigation in the mean-field limit.
This leads to redefining the language shift rates of the AS model through the introduction of a multiplicative Heaviside function term depending on the fraction of speakers.
When population heterogeneity is taken into account,  by replacing the universal threshold by a bell-shaped distribution of thresholds, the step function rate turns into a sigmoid function. 

Depending on interaction rates and initial fractions of speakers, we find three possible outcomes: 
(1) a consensus state characterized by the extinction of one of the two languages; 
(2) a frozen final state identical to the initial one, because of too low a level of interactions (below the thresholds); 
(3) a stable final state where the two languages coexist. 

Our model provides a good fitting of the census data in cases not described by the AS model, e.g., when two languages appear to converge towards coexistence. 
At the same time, it is at least as good as the AS model at fitting the cases where the minority language is on the path to extinction. 

Beyond the AS model, the general approach presented in this letter could be applied to other models of language competition, therefore enabling the inclusion of the effect of the learning process, which is expected to lead to more realistic descriptions of the dynamical scenarios. 



\textit{Acknowledgments.} -- The work was supported by the Estonian Research Council through Grant PRG1059. 
MT acknowledges support from the CUDAN ERA Chair project (EU Horizon 2020 research and innovation program, Grant No. 810961) and is grateful to S.~Maslov for fruitful discussions. The authors also wish to thank K.~Kaski for useful advice and input during the manuscript revision process.

\appendix










\section {Some properties of the AS equation}

We consider the AS equation in the following form  
\begin{equation}
    \frac{dx}{dt} = \left[-(1-\gamma) x(1-x)^a +\gamma x^a (1-x)\right]T^{-1} \, .
\label{AS}
\end{equation}
Here $T$ is the characteristic timescale  and introducing $\tau = t/T$ allows to reduce  the equation to a dimensionless form.  
The parameter $a$ is the volatility; usually it is assumed that $a \geq 1$, but below we also discuss the case of $a < 1$. 
The parameter $\gamma \in [0, 1]$ characterizes the relative prestige of languages; $\gamma = 1/2$ corresponds to full symmetry. 

The terms on the right hand side of Eq.~\eqref{AS} have the meaning of fluxes of individuals from one language to the other. 
Thus, 
\begin{equation}
    r_1 =(1- \gamma) T^{-1} (1-x)^a
    \label{r1}
\end{equation}
is the probability per unit time for a speaker of the language X to switch to the opposite language Y, and 
\begin{equation}
    r_2 = \gamma T^{-1} x^a
    \label{r2}
\end{equation}
is the probability per unit time for a speaker of language Y to switch to language X. 

The stationary point $\bar{x}$ of Eq.~\eqref{AS} is defined by the condition $dx/dt = 0$, leading to
\begin{equation}
    \displaystyle \frac{1-\bar{x}}{\bar{x}} = \displaystyle \left(\frac{\gamma}{1-\gamma}\right)^{1/(a-1)} \, , 
\end{equation}
from which one obtains
\begin{equation} \label{stationary}
 \bar{x} = \displaystyle \left[1+\left(\frac{\gamma}{1-\gamma}\right)^{1/(a-1)}\right]^{-1} \, .
\end{equation}
This fixed point is repulsive for $a > 1$ and attractive for $a < 1$. 

In what follows we assume that the initial condition $x(0)$ satisfies the condition
\begin{equation}
    [x(0)-\bar{x}](a-1) < 0 \, ,
    \label{initial}
\end{equation} 
so that $x(t)$ is a decreasing function of $t$, converging to $x = 0$ for $a > 1$ and to $x = \bar{x}$ for $a < 1$.
This excludes the trivial case $x(0) = \bar{x}$ but  does not lead to any further loss of generality: indeed, AS equation is symmetric under transformation $x \to 1 - x$, $\gamma \to 1 - \gamma$ and this transformation can always be used to satisfy condition \eqref{initial}.

\subsection{Accelerating language decay}

For $a \geq 1$, the AS model predicts that $\ln (x(t))$ is a concave function of $\tau = t/T$, i.e., that its second derivative is negative, 
\begin{equation}
   \frac{d}{d\tau} \left(\frac{ d \ln x}{d\tau} \right) 
   \equiv  \frac{d}{d\tau} \left(\frac{1}{x}\frac{dx}{d\tau}\right)
   < 0 \, .
\label{statement}
\end{equation}
This means that if a language is on the path to extinction, its relative decay rate (fraction of speakers lost per unit time) is always accelerating. 

To prove this, it is enough to show that $|\dot{x}/x|$ is a decreasing function of $x$, i.e., 
\begin{equation}
    f(x) = \frac{d}{dx}  \left(\frac{1}{x}\frac{dx}{d\tau}\right) > 0 \text{  for  } x<\bar{x} \, ,
\end{equation}
because 
\begin{equation}
   \frac{d}{d\tau} \left (\frac{1}{x}\frac{dx}{d\tau}\right) = f(x) \dot{x} \, ,
\end{equation}
and we know that $\dot{x} < 0$ if $x_0 < \bar{x}$ and $a \geq 1$.
Taking the derivative of $f(x)$, using Eq.~\eqref{AS}, and regrouping the terms gives
\begin{equation}
\begin{array}{rll}
    f(x) &= & -(1-\gamma) a (1-x)^{a-1} +     \gamma [(a-1) x^{a-2} (1-x) - x^{a-1}] \medskip \\
    & = & \displaystyle \gamma(a-1)\,x^{\,a-2} + a(1-\gamma) x^{a-1}\left[ \left(\frac{1-x}{x}\right)^{a-1} - \left(\frac{\gamma}{1-\gamma}\right)\right] \, .
\end{array}
\end{equation}
Both terms on the rhs of this expression are positive for $0 < x < \bar{x}$. 
Indeed, the form of the fixed point given by  Eq.~\eqref{stationary} guarantees that the expression in the square brackets is positive for $x < \bar{x}$;  and all the other terms are positive due to the conditions on $a,\gamma, x$. 

Thus, for $a \geq 1$, the function $\ln x(t)$ converges to $-\infty$ for $t \to \infty$ and the corresponding curve is concave.

\subsection{Large time degeneracy}


Here we point out that the large time behavior of the solutions of the AS equation \eqref{AS} is degenerate, in the sense that it depends on the three parameters $T$, $a$, $\gamma$ only through  some combinations thereof, and  in all cases one needs at most two parameters to fit the large time behavior. 
We consider separately the cases $a = 1$, $a > 1$, and $a < 1$.

For $a = 1$ the equation reduces to
\begin{equation}
    \frac{dx}{dt} = -(1 - 2 \gamma) T^{-1} x (1-x) \, , 
\label{AS_time}
\end{equation}
i.e., the degeneracy is exact: not only in the $t \to \infty$ limit but for all $t$ the equation depends only on the combination $T/(1 - 2 \gamma)$.

For $a > 1$ the second term of the equation becomes asymptotically negligible with respect to the first one, so in the limit $t \to \infty$, $x \to 0$ the equation reduces to
\begin{equation}
    \frac{dx}{dt} = -(1 - \gamma) T^{-1} x + O \left(x^{\min(a, 2)} \right) \, .
\label{AS_largeA}
\end{equation}
Therefore, the large time decay is exponential with rate $T/(1 - \gamma)$, while one can determine $a$ and $\gamma$ separately by studying deviations from simple exponential decay at finite values of $x$. 
If $a$ is only marginally larger than 1 the $a$-dependent correction term in \eqref{AS_largeA} is comparable to the linear term up to quite small values of $x$, so determining $a$ is feasible.

Finally, for $a < 1$ the stationary point $\bar{x}$ given in Eq.~\eqref{stationary} is attractive and in its vicinity the equation is approximately linear
\begin{equation}
    \frac{dx}{dt} = -b (x - \bar{x}) + O \left((x - \bar{x})^2 \right)  \, ,
\label{AS_smallA1}
\end{equation}
which also describes an exponential behavior with a rate given by 
\begin{equation}
    b (T,a,\gamma) = T^{-1} (1-a) (1-\gamma) (1-\bar{x})^{a-1} \, .
\label{AS_smallA2}
\end{equation}
Thus, the large time behavior is 
\begin{equation}
    x (t) = \bar{x} + A \exp (-bt) + O (\exp (-2bt)) \, ,
\label{AS_smallA3}
\end{equation}
and the trajectories corresponding to the combinations of parameters $T,a,\gamma$ satisfying
\begin{equation}
\left\{
    \begin{array}{rll}
    \bar{x} (a,\gamma) = \mathrm{const} \medskip \\
    b(T, a , \gamma) = \mathrm{const}
    \end{array}
\right.
\end{equation}
become indistinguishable up to the second order of deviation $x(t) - \bar{x}$. 

In practice, when fitting observational data with the AS equation for $a < 1$, we found that the best fits are indistinguishable from an exponential function and it is impossible to distinguish between solutions with different $a$ from $x(t)$ alone.

That is to say, for all practical purposes the AS equation with any $a < 1$ simply predicts an exponential convergence to language coexistence.

\section{AS model with thresholds}

\subsection{Microscopic derivation of the threshold model} \label{microscopicDer}

The language shift model introduced in this paper is characterized by $X\to Y$ ($Y \to X$) transition rates that depend on thresholds $y^*$ (respectively, $x^*$) on the fractions of speakers of the target language.
Such thresholds can be directly related to the learning process, as discussed in the main text.
We provide here a derivation of such thresholds as a first approximation of the collective dynamics of the many-agent model of bilingualism studied in Ref.~\cite{Scialla-2023a}, which is a generalization of the Minett-Wang model that takes into account the individuals' memory. 
The agents in the model of Ref.~\cite{Scialla-2023a} can be in one of three states, monolingual speakers X and Y and bilinguals (Z). 
However, it is straightforward to simplify it to omit bilinguals.

Following Ref.~\cite{Scialla-2023a}, we measure time in days and assume that an individual interacts daily on average $\Nint$ times with other individuals.
In the hypothesis of well mixed populations, this implies that during a generic day $t$ any individual will interact on average $\Nint \, x(t)$ times with X-speakers and $\Nint \, y(t)$ times with Y-speakers, where $\Nint$ describes phenomenologically the mean daily frequency of social interactions and is determined by various factors such as the average communication habits of an individual, the size and network structure of the community, the density of population, etc. 
The language shift rates are expected to be proportional to the number of encounters with speakers of the target language and can be written as $r_x = J_x \Nint \, x(t)$ and $r_y = J_y \Nint \, y(t)$, where the parameters $J_x$ and $J_y$ are constants (the effective parameters employed in the AS model in the main text are $j_x = J_x \Nint$ and $j_y = J_y \Nint$).
The learning process is taken into account by assuming that a Y-speaker can undergo a Y$\to$X language shift only after interacting with X-speakers at least $K_x^*$ times within the last  $T^*$ days (and analogously for the Y$\to$X language shift). 
Thus, the form of the rates is modified as follows,
\begin{equation}
\begin{aligned}
    \label{Pt}
    & r_x = J_x \Nint x(t) \to 
    J_x \, \Theta\!\left(\Nint \int_{t-T^*}^t x(s) \, ds - K_x^* \right) \, ,
    \\
    & r_y = J_y \Nint y(t) \to 
    J_y \, \Theta\!\left(\Nint \int_{t-T^*}^t y(s) \, ds - K_y^* \right) \, ,
\end{aligned}
\end{equation}
where $\Theta(\cdot)$ is the Heaviside step function, $\int_{t-T^*}^t \Nint x(s) \, ds$ represents total number of encounters with X-speakers in the last $T^*$ days (and analogously with Y-speakers), and for the sake of simplicity the case of neutral volatility $a = 1$ is considered. 
If an X- or a Y-speaker cannot realize at least $K_y^*$ or $K_x^*$ encounters with speakers of the language Y or X, respectively, within the last $T^*$ days, the rates are zero and no language shift can take place -- in this case the fractions of speakers $x(t)$ and $y(t)$ remain unchanged.

If the fractions of X- and Y-speakers vary slowly enough during the time interval $T^*$, one can approximate $\int_{t-T^*}^t x(s) \, ds \approx T^* x(t)$ and $\int_{t-T^*}^t y(s) \, ds \approx T^* y(t)$ and the rates reduce to 
\begin{equation}
\begin{aligned}
    \label{Pt2}
    & r_x \approx J_x \, \Theta\left(x(t) - x^* \right) \, ,
    \\
    & r_y \approx J_y \, \Theta\left(y(t) - y^* \right) \, ,
\end{aligned}
\end{equation}
where we introduced the critical fractions
\begin{equation}
\begin{aligned}
    \label{xymin}
    &x^* = K_x^* / (\Nint T^*) \equiv \nu_x^* / \Nint \, ,
    \\
    &y^* = K_y^* / (\Nint T^*) \equiv \nu_y^* / \Nint \, .
\end{aligned}
\end{equation}
These expressions provide a microscopic interpretation of the critical fractions $x^*$ and $y^*$, where the parameters $\nu_x^* $ and $\nu_y^* $ represent minimal interaction rates. 
In particular, they show that for given minimal frequencies $\nu^*_{x,y}$ the daily interaction rate $\Nint$ plays a crucial role in determining the possibility of a language shift.

\subsection{Sharp learning rates}

The equation of the model with sharp learning rates reads
\begin{equation}
    \frac{dx}{dt} = \left[ -(1-\gamma) x \Theta (1-x-y^*) + \gamma(1-x) \Theta(x - x^*)  \right] T^{-1},
    \label{sharp}
\end{equation}
and includes four parameters  $T$, $\gamma$, $x^*$, and $y^*$. 
The solution of this equation is discussed in detail in the main text. 
Here we recall the possible dynamical regimes and discuss them from a data-fitting angle. 
Similarly to the discussion of the AS equation above, one can, without loss of generality, restrict oneself to considering non-increasing solutions $x(t)$. 
There are then four types of solutions.

0. If $1-y^*<x(0)<x^*$, the state of the system is frozen,  $x(t) = x(0)$.

1. If $x(0) < x^*$ and $x(0)<1-y^*$, the solution is exponentially converging to language extinction, 
\begin{equation}
    x(t) = x(0) \exp(-(1 - \gamma)t/T) \, .
    \label{expdecay}
\end{equation}

2. If $x(0) > x^* > \gamma$ and $x(0) < 1 - y^*$  and $\gamma < 1/2$ the solution still converges to extinction, but consists of two stages: first, 
\begin{equation}
    x(t) = \gamma + (x(0) -\gamma) \exp(-t/T) \, ,
\label{coexistence}
\end{equation} 
until $x(t)$ reaches the value $x = x^*$ at time 
\begin{equation}
t^* = T \ln \frac{x(0) - \gamma}{x^* - \gamma} \, ,
\end{equation}
and, afterwards, continuing towards the extinction state as
\begin{equation}
    x(t) = x^* \exp(-(1 - \gamma)(t - t^*)/T) \, .
\end{equation}

3. Finally, if $1 - y^* > x(0) > \gamma > x^*$ the solution is described by Eq.~\eqref{coexistence} at all times, and converges to language coexistence. 

Other combinations of parameters and initial conditions lead to increasing solutions, which can be reduced to solutions 1-3 by simultaneous replacement $x \to 1 - x$, $\gamma \to 1 - \gamma$.

Notably, the solutions described above also include significant degeneracy: when fitting data to the model one can give but an interval estimate of $y^*$ and, in most cases (except for solution 2), $x^*$. 
In the case of solution 0 it is impossible to determine $\gamma$, $T$ from data, from solution 1 only combination $T/(1 - \gamma)$ is attainable, while from solutions 2 and 3 one can determine $\gamma$ and $T$ separately.

\subsection{Sigmoid learning rates.}

Because people are not identical, but rather heterogeneous in their linguistic traits, such as communication patterns and language learning abilities, due to factors like individual skills, education, and age \cite{Griffiths-2020a,Scialla-2023a},  individuals should be characterized by different threshold values $x^*$, $y^*$.

Assume that the individual thresholds $\xi$ on the fraction $x$ of X-speakers are distributed according to a bell-shaped distribution $w^{-1} f((\xi-x^*)/w)$, where $x^*$ is the mean threshold and $w$ quantifies the width of distribution (and that $f((\xi-x^*)/w)$ is a relatively narrow distribution). Then the step-functions $\Theta(x-x^*)$ in Eq.~\eqref{sharp} change into a corresponding sigmoid function given by 
%
\begin{equation}
    H(x, x^*, w) 
    = \langle \Theta(x - x^*) \rangle 
    = \int_0^\infty d\xi \, \Theta(x - \xi) w^{-1} f((\xi-x^*)/w) 
    = \int_0^x d\xi \, w^{-1} f((\xi-x^*)/w) 
    \equiv F((x-x^*)/w) \, ,
\end{equation}
where $F(x) = \int_0^x dx' \, f(x')$ is the cumulative distribution function corresponding to $f(x)$ and has a sigmoid shape.
%
%
Analogously, the step function $\Theta(1 - x - y^*)$ is replaced by $H(1 - x, y^*, w)$, assuming the same width $w$.


In the absence of hard data, the precise form of the sigmoid function is a matter of choice. To be concrete, we use  
\begin{equation}
    f(x) = C \, \text{cosh}^{-2}(x),
\end{equation}
which is a bell-shaped curve with exponential decay for $|x| \to \infty$. The constant $C(x^*,w)$ is chosen to satisfy the boundary condition
\begin{equation}
    H(1,x^*,w) = \int_0^1 d\xi \, w^{-1} f((\xi-x^*)/w) = 1;   
\end{equation}
leading to
\begin{equation}
\begin{array}{rll}
    H(x,x^*,w) &=& \displaystyle \frac{S(x,x^*,w)-S(0,x^*,w)}{S(1,x^*,w)-S(0,x^*,w)} \, , \medskip \\
    S(x,x^*,w) &= & \{1+ \tanh[(x-x^*)/w]\}/2 \, .
\end{array}
\label{SShape_replacement}
\end{equation}


In Fig.~4 of the main text, the dashed lines represent numerical results for the fraction $x(t)$ for the case of sigmoid learning rates with $w = 0.07$. 
Comparing the solid and dashed lines in Fig.~4(a) of the main text, it can be observed that when the step-function is replaced by the sigmoid function, the neutral equilibrium due to frozen dynamics becomes unstable and the trajectory slowly converges either to $x_\text{I} = 1$ or $x_\text{II} = 0$.

\section{Comparison with observational data}

In this section we provide detailed information about the data sources and the comparison with the numerical simulations of the model introduced.
Table~\ref{table:fits} provides the full list of fitting parameters.

\subsection{Summary of the data used}

\begin{figure}
    \centering
    \includegraphics[width=12cm]{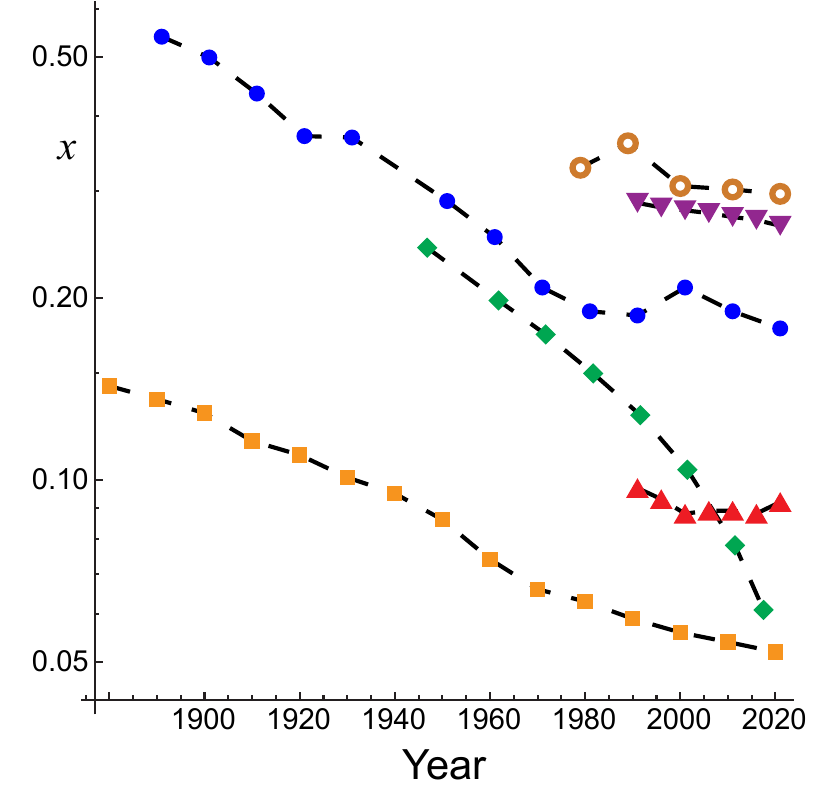}
    \caption{
    Data on the time evolution of the share of minority-language speakers in various countries: Welsh in Wales (blue circles), Swedish in Finland (orange squares), Quechua in Peru (green diamonds), English in Quebec (red triangles), French in the whole of Canada (purple downward triangles), Russian in Estonia (brown empty circles). 
    Connecting dashed lines are guides to the eye.}
    \label{data}
\end{figure}

We collected the following data on prevalence of minority languages in various societies:
\begin{itemize}
    \item Welsh speakers in Wales, according to the UK censuses of 1891 - 2021 \cite{wales}; note that part of this data is used in Ref.~\cite{Abrams-2003a}, although 3 new data points became available since then;
    \item Swedish speakers in Finland in 1880-2020 \cite{finland}; notably, in the middle of this period Finland gained  independence from Russia; however, Swedish has been one of the official languages of Finland since 1863, so the data before and after independence might be reasonably comparable; 
    \item French speakers in Canada according to Canada censuses of 1991-2021 \cite{canada} (we were not able to find earlier data); since Canada has a substantial part of the population whose native language is neither English nor French, we use here the ratio of French-speakers to the sum of French- and English-speakers as the $x$ variable; 
    \item Similar data for English-speakers in Quebec \cite{canada};
    \item Data on fraction of Russian-speakers in Estonia\cite{estonia} in the censuses of 2001-2021; note that, contrary to the case of Swedish in Finland, the status of Russian language changed drastically at the collapse of the USSR in 1991, so that earlier data, although shown in \fig{data}, are not comparable;
    \item native Quechua speakers in Peru according to 2021 census \cite{quechua}; here, following Ref.~\cite{Abrams-2003a}, we use the data of fractions of native Quechua-speakers in different age cohorts as a proxy for how prevalence of Quechua language changed in time; years for this dataset correspond to the weighted average years of birth of corresponding age cohorts.
\end{itemize}

There are several important caveats concerning these data. 
First, one has to be aware that slight changes in the exact wording of census questions (e.g., ``what language you use most often'' vs ``what language you identify with'', etc.) can lead to changes in the reported numbers. 
Second, the Quechua dataset relies on self-reporting of the first language the respondents learned in their childhood; so, like any non-synchronous self-reporting data, it might be prone to biases and misrepresentation. 
Third, there is a question of bilingualism, which is wide-spread in all six cases. 
The way these bilinguals are treated is slightly different in different datasets: in Welsh dataset everybody who claims to use Welsh language is classified as Welsh-speaker regardless of their proficiency in English; in Peruvian dataset people are classified according to the first language they learned in their childhood; in the other four cases classification is based on the primary language or language used at home.

%

\subsection{Semi-frozen regimes}

Figure~\ref{data} summarizes the data for all these six systems. 
Qualitatively, one clearly sees two types of behavior: while for Wales, Finland and Peru the inverse average slope is between 50 and 130 years, it is 370 years for Canada, 560 for Quebec and 700 for Estonia (where we use the fit based on the last 3 points because the previous two ones are not comparable, as explained above). 

The smaller timescale, which is comparable with human lifetime, is what one would expect for a characteristic time of switching between languages. Also, if one interprets Welsh and Finnish data as converging to a long-time equilibrium (see below), the corresponding relaxation times for Wales, Finland and Peru turn out to be even closer, lying in the range of 27 to 55 years.

The language shift in the three other cases is happening on a completely different timescale. In fact, it is not clear whether it is happening at all: the slow change in language composition of Canada, Quebec and Estonia might possibly be explained by external migration and/or discrepancy in the rates of population growth of the two communities rather than by actual switching from one language to another.
Note that this slow dynamics happens despite the fact that communities are definitely interacting with each other: indeed, almost half of Estonian population lives in the mixed-language environment of the Tallinn metropolitan area and is exposed to both Estonian and Russian on a daily basis; the same is true for Canada where both English and French are official languages. 
We suggest that in these cases the interaction between different linguistic communities, although clearly present, is not strong enough to make reaching proficiency in the second language cost-effective for the large majority of the population. 
In the simple piecewise-linear model this corresponds to a completely frozen regime (regime 0 described in Sec.~\ref{microscopicDer} above), and in the case of sigmoid learning rates -- to the slow dynamics shown in yellow, green and teal in Fig.~4(a) of the main text.

\subsection{Decaying regimes}

%
\begin{figure}
    \centering
    \includegraphics[width=17cm]{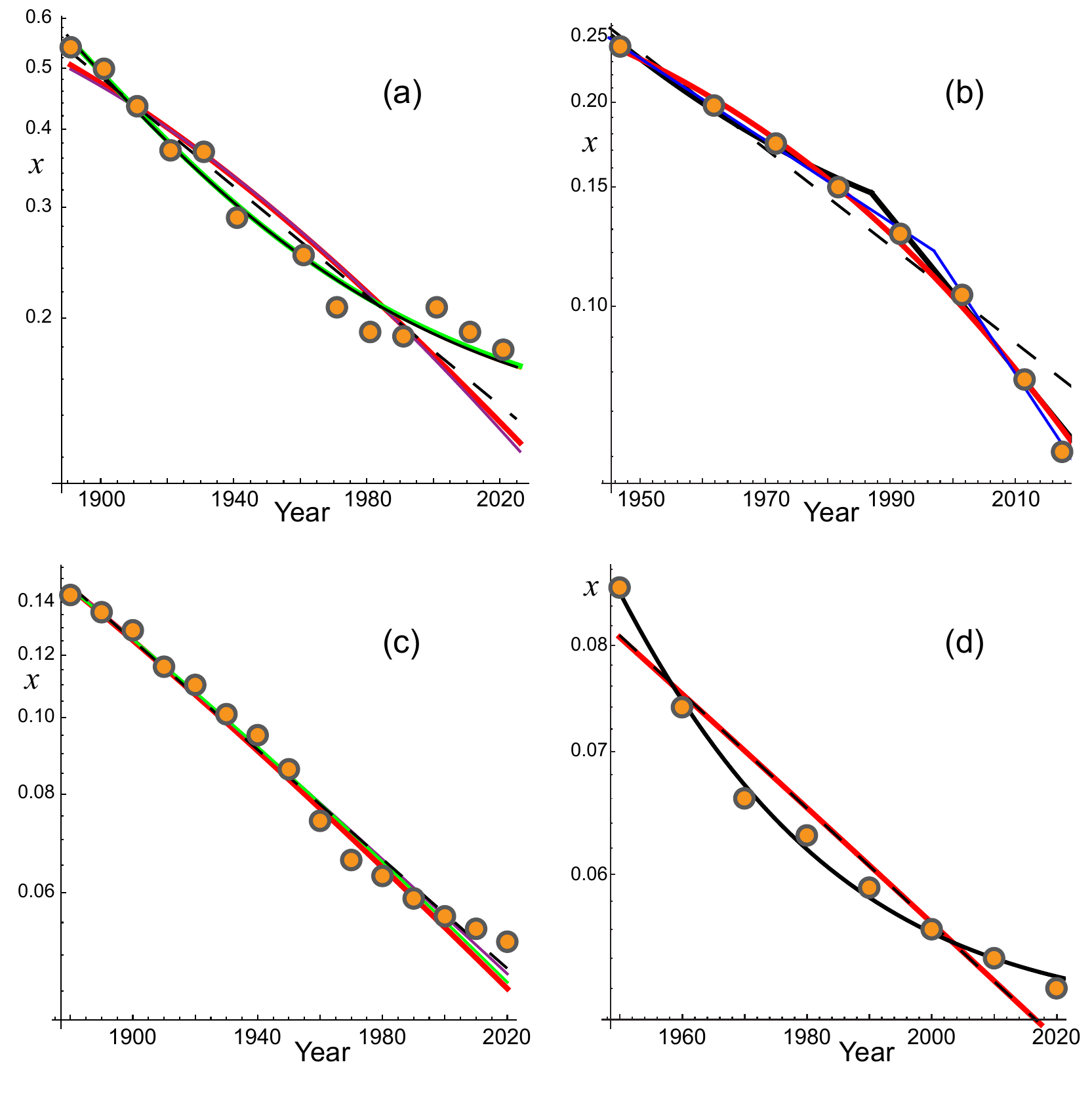}
    \caption{
    Data on the prevalence of decaying languages compared to fits by various models: (a) Welsh in Wales, (b) Quechua in Peru, (c) and (d) Swedish in Finland. 
    In all panels, the dashed black line is a simple exponential fit $x(t) = x_0 \exp(-t/t_0)$ and the red line is the best fit obtained using the AS equation with $a > 1$. 
    Solid black lines correspond to best fits made with the piecewise-linear model introduced in this paper, regime 3 (see section IIB above) in case of (a) and (d) and regime 2 in case of (b); the best fit in case (c) corresponds to regime 1 and thus coincides with the dashed black line. 
    The blue line in panel (b) corresponds to a bi-exponential fit \eq{biexponential} and the purple and green lines correspond to auxiliary fits made by using the AS equation with additional constraints, see the text and Table~\ref{table:fits} for details and parameters values.}
    \label{datafit}
\end{figure}

In three other cases there is a clear decay in the prevalence of the minority language.

Consider the prevalence of Welsh in Wales first. 
The dashed black line in Fig.~\ref{datafit}(a) shows the best fit with exponentially decaying function $x = x_0 \exp(-t/T)$. 
The red and purple lines are best fits with the Abrams-Strogatz equation with $a = 1$ (which corresponds to the best fit for any $a \geq 1$) and $a = 1.3$, respectively. 
Clearly, these fits fail to describe the slow-down in the decay of Welsh observed in recent decades, and are, in fact, worse than a simple exponential fit. 

In turn, the black line is an exponential fit of the form \eqref{coexistence}, which shows a better agreement with data. 
Note that, as discussed above, the AS equation with $a<1$ is, in the first order, equivalent to \eqref{coexistence}. 
At the same time, the AS equation has an additional free parameter, so one could have expected that it is possible to achieve a better fit to the data by adjusting the second-order corrections. 
However, the fits with the AS equation with various $a < 1$ are essentially indistinguishable from simple exponential. The best one we were able to find is shown with a green line in Fig.~\ref{datafit}(a). It corresponds to $a = 0.09$ (i.e., way beyond the usually assumed range $a = 1.3 \pm 0.35$) and has a sum of residuals just 1\% smaller than the simple exponential fit (see Table~\ref{table:fits} for the values  of the fitting parameters). 

\begin{table}[ht!]
\centering
\begin{tabular}{||l |l | l | l | l||} 
\hline
\hline
Line& Fitting function & Parameters & Initial & $10^4 \langle \delta_i^2 \rangle$\\
& & & condition& \\
\hline
\multicolumn{5}{||l||}{Welsh in Wales}  \\
\hline 
Dashed black &$x_0 \exp (-t/T)$ & $ T = 100$ years & $x_0 = 0.530$ & 5.4 \\
Black & Regime 3 & $\gamma = 0.13$, $T= 55$ years & $x_0 = 0.556$ & 2.71 \\
Red & AS with $a\geq 1$ &$a=1,$ & $x_0 = 0.506$ & 8.9 \\
& & $T/(1-2\gamma)= 69$ years& & \\
Purple & AS with $a = 1.3$ &$\gamma = 0, T= 62$ years & $x_0 = 0.498$ & 10.3 \\
Green & AS with $a <1$ &$ a= 0.086,\, \gamma = 0.15,$ & $x_0 = 0.556$ & 2.68 \\
& & $T= 50$ years& & \\
\hline
\multicolumn{5}{||l||}{Quechua in Peru}  \\
\hline 
Dashed black &$x_0 \exp (-t/T)$ & $ T = 61$ years & $x_0 = 0.251$ & 0.82 \\
Black & Regime 2 & $\gamma = 0.103$, $x^* = 0.147$& $x_0 = 0.244$ & 0.106 \\
& & $T= 34.5$ years& & \\
Red & AS with $a\geq 1$ &$a=1.4, \gamma = 0.534$ & $x_0 = 0.24$ & 0.11 \\
& & $T= 9$ years& & \\
Blue & Eq. \eqref{biexponential} &$x^* = 0.121$, & $x_0 = 0.243$ & 0.024 \\
& & $T_1= 83$, $T_2 =31$ years& & \\
\hline
\multicolumn{5}{||l||}{Swedish in Finland, 1880-2020}  \\
\hline 
Dashed black &$x_0 \exp (-t/T)$ & $ T = 125$ years & $x_0 = 0.147$ & 0.096 \\
Red & AS with $a\geq 1$ &$a=1.6, \gamma = 0.$ & $x_0 = 0.146$ & 0.090 \\
& & $T = 106$ years& & \\
Purple &AS with $a=1$ & $T/(1-2\gamma)= 113$ years & $x_0 = 0.1465$ & 0.091 \\
Green &AS with $a=2$ & $\gamma = 0,T= 102$ years & $x_0 = 0.1456$ & 0.091 \\
\hline
\multicolumn{5}{||l||}{Swedish in Finland, 1950-2020}  \\
\hline 
Dashed black &$x_0 \exp (-t/T)$ & $ T = 138$ years & $x_0 = 0.081$ & 0.08 \\
Black &Regime 3 & $\gamma = 0.050,$ & $x_0 = 0.0855$ & 0.005 \\
& & $T= 27.5$ years& & \\
Red & AS with $a\geq 1$ &$a=1, \gamma = 0.308$, & $x_0 = 0.0808$ & 0.083 \\
& & $T/(1-2\gamma)= 130$ years& & \\
\hline
\hline
\end{tabular}
\caption{Parameters of fits shown in \fig{datafit}.}
\label{table:fits}
\end{table}

Consider now the case of Quechua in Peru. 
The current trajectory [see \fig{datafit}(b)] clearly points to the eventual extinction of Quechua, and the decay is accelerating. 
This is generally the behavior predicted by the AS equation, so it is no surprise that it is able to fit the data reasonably well (see the red curve). 
Interestingly, according to the parameters of the fit, the accelerating decay of Quechua is mostly due to an almost complete stop of the Spanish to Quechua language shift. Indeed, compare the state of the system at $x = 0.25$ (circa 1950) to that at $x = 0.06$ (circa 2020). While the rate $r_1$ \eqref{r1} of Quechua to Spanish conversion increases only marginally between these points, from $(29 \text{ years})^{-1}$ to $(21 \text{ years})^{-1}$, the rate of $r_2$ \eqref{r2} of inverse conversion  collapses from  $(120 \text{ years})^{-1}$  to $(870 \text{ years})^{-1}$.

The data for Quechua in Peru can be fit with the regime 2 of the piecewise-linear model, as described in Appendix B2, see black line in Fig.~\ref{datafit}(b). 
It is noteworthy that, although in regime 2 the shape of the trajectory $x(t)$ has a kink, which is due to a piecewise-linear character of the model, it fits the data as well as the AS fit (formally speaking, even slightly better). 
The qualitative underlying picture suggested by this fit is that the difference between the two stages is due to an abrupt stop of the Spanish to Quechua language shift at a critical fraction of Quechua speakers $x^* \approx 0.147$. 

Interestingly, the same data might can be fit even better by a simple combination of two exponents:
\begin{equation}
    x(t) = \left\{
    \begin{array}{ll}
    x_1 \exp(-t/T_1) &    \text{for }  x\geq x^* = (x_1^{T_1} x_2^{-T_2})^{1/(T_1-T_2)}\medskip \\
    x_2 \exp(-t/T_2) &  \text{for }  x< x^*
    \end{array}
    \right.
    \label{biexponential}
\end{equation}
where $T_1>T_2$ is implied and $x_1$ coincides with initial condition $x_0=x(t=0)$. However, this is no more than a curious observation because at the moment we are not able to suggest any microscopic mechanism, which could produce dynamics \eqref{biexponential}.

Finally, consider the fate of Swedish language in Finland. This is the longest dataset available, spanning almost 150 years. Considering this dataset as a whole, it is well approximated with simple exponential decay \eqref{expdecay}, see dashed line in Fig.~\ref{datafit}(c). 
Formally speaking, the AS model with $a \geq 1$ gets a tiny improvement in the quality of the fit in exchange for additional fitting parameters. 
However, it is notable that AS is even worse at describing the slow-down in the decay of Swedish observed in recent decades.

This observational dataset, however, spans multiple important events in the history of Finland, including, among other things, the separation of Finland from Russia and several wars. 
It is quite possible that assuming a constant environment (i.e., constant values of the model parameters) throughout this period is too crude an approximation. 
It is also clear from the data that the shape of $x(t)$ undergoes a significant change around the time of the Second World War.
Thus, we decided to consider the post-war period of the Finnish data as a separate dataset, shown in Fig.~\ref{datafit}(c). 
It is clear that this regime fits almost perfectly with exponential convergence to language coexistence (regime 3 as explained in Sec. IIB, solid black line), while the AS equation with $a \geq 1$ (red line) does not give any improvement compared to a simple exponential fit (black dashed line).



\end{document}